\documentclass{jetpl}
\usepackage{epsfig,amssymb,amsfonts,bm}
\twocolumn

\DeclareMathSymbol{\varSigma}{\mathord}{letters}{"06}
\newcommand{\be}{\begin{equation}}
\newcommand{\ee}{\end{equation}}

\newcommand{\vp}{\varphi}
\newcommand{\x}{{\bm x}}

\newcommand{\p}{{\bm p}}
\newcommand{\bea}{\begin{eqnarray}}
\newcommand{\eea}{\end{eqnarray}}

\title{Dark quark domains}
\rtitle{Dark quark domains}

\author{D.\,V.\,Antonov}
\address{Centro de F\'\i sica das Interac\c c\~oes
Fundamentais (CFIF), Departamento de F\'\i sica, Instituto Superior
T\'ecnico, Av. Rovisco Pais, 1049-001,
Universidade T\'ecnica de Lisboa, Lisbon, Portugal}
\author{A.\,V.\,Nefediev}
\address{Institute of Theoretical and Experimental Physics,
117218, B.Cheremushkinskaya 25, Moscow, Russia}
\author{J.\,E.\,F.\,T.\,Ribeiro}
\address{Centro de F\'\i sica das Interac\c c\~oes
Fundamentais (CFIF), Departamento de F\'\i sica, Instituto Superior
T\'ecnico, Av. Rovisco Pais, 1049-001,
Universidade T\'ecnica de Lisboa, Lisbon, Portugal}
\rauthor{D.\,V.\,Antonov, A.\,V.\,Nefediev, J.\,E.\,F.\,T.\,Ribeiro}

\abstract{Formation of stable domains filled with strongly correlated coherent
quark matter is discussed in general terms and is exemplified further in the framework of the Generalised Nambu--Jona-Lasinio model. It is argued that such domains, if exist in the Universe, appear dark to
an external observer.}

\PACS{12.38.Aw 12.38.Lg}

\begin{document}
\maketitle

\section{Introduction}

The QCD vacuum is known to be a highly nontrivial medium consisting of fluctuations of gluonic fields, various condensates, and so on. 
One of the best studied and microscopically understood phenomena related to these nontrivial properties of the vacuum is the spontaneous breaking 
of chiral symmetry --- a global symmetry inherent to the QCD Hamiltonian, which, however, is not respected by its lowest eigenstate, 
that is by the vacuum. As a result, the entire tower of hadronic states built on top of the vacuum lacks this kind of symmetry as well. 
While the effect of spontaneous chiral symmetry breaking is expected to play no role for highly excited states in the spectrum of hadrons --- 
the so-called effective chiral restoration for highly excited hadrons (see review \cite{G6} and references therein)
--- the spectrum of low-lying states is strongly affected by chiral symmetry 
breaking. The most prominent example of the manifestation of chiral symmetry breaking in the spectrum of hadrons is provided by the pion, which plays the role of the pseudo-Goldstone boson associated with chiral symmetry breaking.
Another 
manifestation of chiral symmetry breaking is given by the constituent mass, of the order of a few hundred 
MeV, which a light quark acquires in the course of its propagation through the vacuum.

The phenomenon of chiral symmetry breaking is a
highly nonlinear effect, 
which is described in terms of nonlinear gap equations, often referred to as mass-gap equations. Such equations are known to 
possess at least two solutions --- the trivial solution for the chirally symmetric vacuum $|0\rangle_0$ and a nontrivial solution describing the 
physical chirally nonsymmetric vacuum $|0\rangle$. In this Letter, we discuss in general terms a physical picture which can arise from a possible 
existence of further nontrivial solutions to the mass-gap equation --- the so-called replicas. 
The existence of such solutions was emphasised many times in the framework of the Generalised Nambu--Jona-Lasinio model (GNJL). In particular, this was done in Ref.~\cite{rep1} for the
quadratic confining potential, in Ref.~\cite{rep2,rep25} for the linear potential, and
in Ref.~\cite{rep3} for an arbitrary power-like confining potential. In Ref.~\cite{rep4}, similar solutions were
considered in different approaches, while general arguments in favour of the existence of replicas were given in a recent
paper \cite{rep5}.
For the sake of simplicity, we will assume 
throughout the present paper that only one such excited solution to the mass-gap equation exists, which will be denoted as $|R\rangle$. Once this is the case, the corresponding state
possesses the energy density higher than that of the vacuum $|0\rangle$. However, since chiral symmetry is broken in the state $|R\rangle$, the latter is still 
energetically more preferable than the trivial, 
chirally-symmetric, vacuum $|0\rangle_0$. Below, it is assumed that such a replica state 
fills a finite-size domain in the Universe, and the problem of stability of such a domain is addressed. We start with a 
qualitative model-independent consideration, exemplifying it further by explicit calculations in the GNJL 
framework.

A replica-filled domain at issue is described by just two parameters --- its radius $R$ (the volume being
denoted as $V$) and the momentum scale $\Lambda_\chi$ determining
the characteristic distance $R_\chi$ at which chiral
symmetry breaking occurs: $R_\chi\simeq 1/\Lambda_\chi\sim 1\div 1.5$~fm.
Indeed, it is legitimate to define $\Lambda_\chi$ in this way because the 
impact of chiral symmetry breaking is significantly 
weaker inside the domain than in the outer space, which is filled 
with the unexcited vacuum. For this reason, we can identify  
$\Lambda_\chi$ with a typical momentum scale of chiral symmetry breaking in the true QCD vacuum. 
As such, its value can be estimated as  
$250\div 300$~MeV. Accordingly, the replica energy density relative to that of the unexcited vacuum appears 
as $\varepsilon=(\Lambda_\chi)^4>0$.

Below we impose constraints on the values of $R$, which ensure stability of the replica-filled domain.

\section{A microscopic content of the excited vacuum state}

In the true QCD vacuum, 
spontaneous breaking of chiral symmetry leads to the formation of the
chiral condensate --- 
a coherent-like state of strongly correlated $^3P_0$ quark--antiquark pairs. Being yet another genuine solution to the
mass-gap equation, the excited state $|R\rangle$ admits a similar microscopic interpretation.
However, the way quark--antiquark pairs there get correlated to form the condensate 
is different --- while the quantum numbers $^3P_0$ are obviously the same, in order to ensure the state $|R\rangle$ to be a scalar, the radial wave 
functions of such pairs appear to be excited. Clearly, while in the infinite-volume limit, the $|0\rangle$ and the $|R\rangle$ states are strictly orthogonal, for finite volumes $V$ 
their overlap is nonvanishing and reads:
\be
\langle 0|R\rangle=e^{-\Lambda_\chi^3 V}.
\label{overlap}
\ee
Below, by means of the explicit 
expressions found in the GNJL model, we represent 
Eq.~(\ref{overlap}) in terms of the chiral angle characterising a given replica $|R\rangle$. 

Thus, 
we conclude that, from the viewpoint of an external observer, the replica-filled domain is nothing but a 
localised in space coherent cloud of strongly correlated quark--antiquark pairs.

\section{Decays of the replica state and the minimal size of the domain}

We are now in a position to estimate the minimal possible size of a stable replica-filled domain.
Indeed, Eq.~(\ref{overlap}) suggests that the overlap between the replica and the vacuum increases as the volume of the domain decreases. 
Accordingly, the lifetime of the domain decreases fast with the decrease of its volume. To quantify this
effect, one can use the Schr{\"o}dinger equation for the domain of volume $V$,
\be
i\frac{\partial}{\partial t}|R\rangle=\hat{\cal H}|R\rangle,
\label{seq1}
\ee
where $\hat{\cal H}$ is the QCD Hamiltonian. Namely, we notice that: (i) if the excited state $|R\rangle$ decays
into the unexcited one, $|0\rangle$, during the time $\tau$, the derivative on the left-hand side of Eq.~(\ref{seq1}) can be
approximately substituted by a finite difference, $\partial |R\rangle/\partial t\to (|0\rangle-|R\rangle)/\tau$; (ii)
for a sufficiently large volume $V$, $|R\rangle$ is an approximate eigenstate of the Hamiltonian with the
eigenenergy $V\varepsilon$. Temperature-dependent hadronic contributions to the energy density \cite{agas} can be shown to be negligible as compared to $\varepsilon$, up to temperatures $\sim 25$~MeV \cite{rep5}.

Finally, projecting Eq.~(\ref{seq1}) onto the vacuum state $|0\rangle$, using Eq.~(\ref{overlap})
and the fact that this state is normalised as $\langle 0|0\rangle=1$, one arrives at the following estimate:
\be
\tau^{-1}\sim V\varepsilon e^{-\Lambda_\chi^3 V}\sim V\Lambda_\chi^4 e^{-\Lambda_\chi^3 V}.
\label{w}
\ee
Thus, we find that the lifetime of the replica-filled domain can be estimated as 
\be
\tau\sim\frac{1}{\Lambda_\chi}\left[\frac{e^{\Lambda_\chi^3 V}}{\Lambda_\chi^3 V}\right].
\ee
The condition of stability is obviously formulated as the requirement that
$\tau$ is larger than the age of the Universe, $\tau>T_U$, that is
\be
\frac{e^{\Lambda_\chi^3 V}}{\Lambda_\chi^3 V}>\Lambda_\chi T_U,
\label{eq}
\ee
which yields an estimate for the volume of the domain in terms of 
the dimensionless parameter $\Lambda_\chi T_U$.

For $\Lambda_\chi T_U\gg 1$, relation (\ref{eq}) can be approximately resolved with respect to the volume as
\be
V>\Lambda_\chi^{-3}\ln\left[\Lambda_\chi T_U\ln(\Lambda_\chi T_U)\right],
\ee
which gives for the minimal radius of the domain the value
\be
R_{\rm min}=\Lambda_\chi^{-1}\left(\ln\left[\Lambda_\chi T_U\ln(\Lambda_\chi T_U)\right]\right)^{1/3}
\label{RR0}
\ee
or, numerically,
\be
R_{\rm min}\simeq 5~\mbox{fm}.
\label{eq2}
\ee
Notice that, because of the weak logarithmic, further weakened by the cubic root, dependence 
of the result (\ref{RR0}) on $\Lambda_\chi T_U$, the estimate (\ref{eq2}) is very stable against
variations of the values of both $\Lambda_\chi$ and $T_U$.
The obtained estimate for $R_{\rm min}$ is of the order of a
typical nucleus size.

\section{The maximum size of the domain compatible with its gravitational stability}

In this section, we estimate the maximal possible size of the replica-filled domain. The existence of such an upper bound for the domain size is a necessary condition for the 
stability of the domain against the gravitational collapse \cite{rep5}. Indeed, for a constant energy density $\varepsilon$ at issue, the mass of a
replica-filled domain increases linearly with its volume. Therefore, the Schwarzschild radius 
$r_g=2G\cdot (V\varepsilon)$, where $G$ is the gravitational constant, increases with the size of the domain as 
\be
r_g\propto R^3.
\ee
The gravitational stability condition requires that $r_g<R$, which leads to an estimate
\be
R_{\rm max}\sim 1/\sqrt{\varepsilon G}.
\ee
For a more rigorous analysis, one can 
apply the Tolman--Oppenheimer--Volkoff (TOV) equation for the pressure inside 
a spherically-symmetric domain \cite{rep5}, and 
use its well-known solution for the case of a constant energy density \cite{TOV}. It yields a more 
restrictive upper bound
\be
\label{radius}
R_{\rm max}=1/\sqrt{3\pi G\Lambda_\chi^4}\sim 10{\,}{\rm km}.
\ee
Notice that the obtained estimate for the maximum radius is of the order of a
typical radius of a neutron star, which is also about 10~km.

A comment is in order here on a possible back reaction of the gravitational field of the domain 
on the equation of state (EoS) $\varepsilon=(\Lambda_\chi)^4$. Indeed, solving the TOV equation 
together with the EoS, one finds a solution which defines unambiguously the gravitational metric inside the domain 
(cf. Ref. \cite{rep5}). The chiral condensate in such a curved space differs, generally speaking, from its flat-space counterpart, whose value was initially used in the EoS. Correcting the EoS accordingly and 
plugging it again into the TOV equation, one arrives at an iterative procedure. Fortunately, it turns out 
that, owing to the numerical smallness of the corresponding gravitational correction to the chiral condensate, the iterations terminate already at the zeroth one, that is the EoS 
and the mass-gap equation decouple from each other. Namely, the corrected chiral condensate 
has the form \cite{rep5}:
\be
\langle\bar\psi\psi\rangle=\langle\bar\psi\psi\rangle_0 \left[1+{\cal O}((R_\chi/R)^2)\right].
\ee
This result shows that, since chiral symmetry breaking and gravity are effective at different length scales,
the equations describing them can be clearly separated. In particular, one 
can see that, even for a domain of the smallest possible size (\ref{RR0}), the impact of its gravitational field on the chiral condensate is negligible. 

\section{Replicas within the GNJL}

One of the most successful models in studies of various phenomena related to chiral symmetry breaking is the famous
Nambu--Jona-Lasinio (NJL) model \cite{NJL}. This model still possesses a few shortcomings, among which the most essential ones are the
absence of an intrinsic scale, that shows up in the ultraviolet regularisation of the loop integrals, and the
absence of confinement. A generalisation of the NJL model was suggested in Refs.~\cite{rep1,Orsay,Lisbon}
with the purpose to include confinement, that resulted in the proposition of the GNJL. 
Finally, the direct connection between the $^3P_0$ quark content of 
the vacuum --- in terms of the chiral angle
$\vp_{\bm p}$ (see below) --- and the mechanism for the spontaneous breaking of chiral symmetry was shown explicitly in Ref.~\cite{Lisbon}. 

The Hamiltonian of the GNJL reads:
\begin{eqnarray}
H&=&\int d^3 x\bar{\psi}({\bm x},t)\left(-i{\bm \gamma}\cdot
{\bm \bigtriangledown}+m\right)\psi({\bm x},t)\nonumber\\
&+&\frac12\int
d^3 xd^3y\;J^a_\mu({\bm x},t)K_{\mu\nu}({\bm x}-{\bm y})J^a_\nu({\bm y},t),
\label{H}
\end{eqnarray}
where $J_{\mu}^a({\bm x},t)=\bar{\psi}({\bm x},t)\gamma_\mu\frac{\lambda^a}{2}\psi({\bm x},t)$, and 
$\psi({\bm x},t)$ is the quark field. Thus, the
interquark interaction is parametrised by the quark kernel $K_{\mu\nu}$ which contains confinement explicitly, in
the form of a rising potential. The latter can be supplied with various extra terms, for instance the perturbative
Coulombic one. The confining interaction between quarks brings a
nonperturbative scale into the problem and models radial excitations of ${}^3P_0$ pairs mentioned above,
in Sect.~2. Chiral symmetry breaking can then be described by the summation of 
loop diagrams for valence quarks (which leads to the mass--gap equation), while mesons are obtained from the Bethe--Salpeter
equation for the quark--antiquark bound states. Besides that, GNJL is known to fulfil the low-energy theorems
of Gell-Mann, Oakes, and Renner \cite{GOR}, Goldberger and Treiman \cite{GT}, Adler self-consistency zero \cite{ASC}, 
the Weinberg theorem \cite{Wein}, and several others. We would like to emphasise a universal nature
of the above low-energy theorems, regardless of a particular form of gluonic interactions which result 
in a chiral-symmetry-breaking quark kernel. 
Consequently, irrespectively of a
particular form of the {\it confining} kernel $K_{\mu\nu}$, GNJL provides a reliable phenomenological approach to chiral symmetry
breaking, which incorporates effects of confinement. Thus, owing to the fact that neither 
qualitative nor quantitative predictions of the GNJL depend 
on a particular form of 
$K_{\mu\nu}$, it suffices to impose this kernel to be confining
and to introduce the scale of chiral symmetry breaking, $\Lambda_\chi$.

Furthermore, all possible quark--quark interactions described by the
Hamiltonian (\ref{H}) include quark self-interactions.
These self-interactions can be eliminated by the use of an
appropriate Bogoliubov--Valatin transformation from bare quarks to the dressed ones.
Such a transformation can be conveniently parametrised by means of the so-called chiral angle
$\vp_{\bm p}$ (${\bm p}$ being
the relative momentum of a dressed quark--antiquark
pair) \cite{Orsay,Lisbon}:
$$
\psi^\alpha(\x)=\sum_{\p,s}e^{i\p\x}
[b^\alpha_{\p s}u_s(\p)+d^{\alpha\dagger}_{\p s} v_s(-\p)],
\label{psi}
$$
\begin{eqnarray}
u(\p)&=&\frac{1}{\sqrt{2}}\left[\sqrt{1+\sin\vp_\p}+
({\bm\alpha}\hat{\p})\sqrt{1-\sin\vp_\p}\right]u_0(\p),\nonumber\\
v(-\p)&=&\frac{1}{\sqrt{2}}\left[\sqrt{1+\sin\vp_\p}-
(\bm{\alpha}\hat{\p})\sqrt{1-\sin\vp_\p}\right]v_0(-\p)\nonumber,
\end{eqnarray}
where $\alpha$ is the colour index of $N_C$ colours. It is
convenient to define the chiral angle varying in the range
$-\pi/2<\vp_\p\leqslant\pi/2$ and respecting the boundary
conditions:
\be
\vp_\p(\p=0)=\pi/2,\quad \vp_\p(|\p|\to\infty)\to 0.
\label{bc}
\ee

The normal-ordered Hamiltonian (\ref{H}) takes the form:
\be
H=E_{\rm vac}+:H_2:+:H_4:,
\label{H3}
\ee
and the usual procedure to minimise the vacuum
energy $E_{\rm vac}$ is to set the quadratic part
$:H_2:$ to be diagonal. Then the
corresponding mass-gap equation,
\be
\delta E_{\rm vac}[\vp]/\delta\vp_\p=0,\quad
E_{\rm vac}[\vp]=\langle 0[\vp]|H|0[\vp]\rangle,
\label{mge}
\ee
ensures the anomalous Bogoliubov
terms $b^\dagger d^\dagger$ and $db$ to be absent in
$:H_2:$. The mass-gap equation (\ref{mge}) is known to possess one trivial solution $\vp_\p\equiv 0$ 
and multiple
nontrivial solutions \cite{rep1,rep2,rep3,rep4,rep5}. For the sake of simplicity, let us assume 
the existence of just
two nontrivial solutions --- one describing the vacuum and the other defining one replica (more than one replica 
will not change the argument).

As soon as the mass-gap equation is solved and a nontrivial chiral
angle is found, the Hamiltonian (\ref{H3}) takes a diagonal
form,
\be
H=E_{\rm vac}+\sum_{\p,\alpha,s}
E_\p[b^{\alpha\dagger}_{\p s}b^\alpha_{\p s}+d^{\alpha\dagger}_{\p s}d^\alpha_{\p s}]+\ldots,
\label{H2diag}
\ee
where $E_\p$ is the dressed-quark dispersive law, and the ellipsis stands for the omitted terms responsible for the formation of bound states of quarks. A Fock space can be built on top of the
nontrivial vacuum $|0\rangle$ upon the action of the quark creation operators.

As it always happens after a Bogoliubov--Valatin transformation, the
new vacuum contains an infinite set of
strongly correlated $^3P_0$ quark-antiquark pairs \cite{Lisbon}. The same holds true for the replica $|R\rangle$, so
it is not hard to find explicitly the operator which defines a pseudounitary transformation from $|0\rangle$ to
$|R\rangle$:
\be
|R\rangle=e^{Q-Q^\dagger}|0\rangle,\quad Q^\dagger=\frac12\sum_{\p}\Delta\vp_\p C_\p^\dagger,
\label{S0}
\ee
where $C_\p^\dagger=b^{\alpha\dagger}_{\p s}[({\bm \sigma}\hat{{\bm
p}})i\sigma_2]_{ss'}d^{\alpha\dagger}_{\p s'}$, with
$\sigma$'s being the $2\times 2$ Pauli matrices, and
$\Delta\vp_\p$ being the difference between the vacuum and the replica chiral angles.
The operator $C_\p^\dagger$ creates a $^3P_0$ quark-antiquark pair with zero
total momentum and the relative three-momentum $2{\bm p}$, while the operator $\exp[Q-Q^\dagger]$ 
creates a strongly correlated cloud of such pairs. 

Using the commutation relations for the quark operators $b$ and $d$,
one can readily arrive at Eq.~(\ref{overlap}) in the form:
\be
\langle 0|R\rangle=\exp\left[V\int\frac{d^3p}{(2\pi)^3}\ln\left(\cos^2\frac{\Delta\vp_\p}{2}\right)\right].
\label{qq12}
\ee 
The function $\Delta\vp_\p$ is predominantly nonvanishing
at $|\p|\lesssim\Lambda_\chi$, while decreasing fast at larger relative momenta. For this reason, 
the integral in the exponent of Eq.~(\ref{qq12}) is ${\cal O}(\Lambda_\chi^3)$, and only an overall numerical coefficient of the order of unity depends on the details of the interquark interaction.

This way, the scale of chiral symmetry breaking $\Lambda_\chi$, initially introduced through the quark 
kernel $K_{\mu\nu}$, appears naturally in the second-quantised formalism.
Thus we have obtained a particular realisation of the physical picture which was described in the previous sections in general terms. 

\section{Discussion}

In this Letter, we have discussed some physical consequences of having 
more than one realisation of chiral symmetry breaking in QCD. Indeed, given a highly nonlinear nature of the phenomenon of
chiral symmetry breaking, the possibility for the mass-gap equation to possess more than one nontrivial solution does
not look unnatural --- on the contrary, one would need to specially arrange for a nonlinear equation to possess 
just one solution. In the literature, 
the presence of multiple solutions to the mass-gap equation was verified in a
number of NJL-type models, and the corresponding excited vacuum states were constructed explicitly. 
Among such states, only the lowest one can be associated with the genuine vacuum of the theory, while 
the others should
be interpreted as some scalar excitations on top of it --- we use the term replicas for them. In
particular, by putting forward the conjecture that replicas may fill domains in the Universe, we have further explored stability
conditions for the size of such domains. It turns out that the allowed domain radius can vary from just a few fermi up
to a few kilometres. 

The only way for the domain to be detected by an external observer is through its decay to the 
QCD vacuum, with the release of the vacuum energy in the form of a cloud of hadrons (predominantly pions), which
would further annihilate, producing light. Notice however that, due to the very strong correlations between 
quarks inside the domain, which result in the exponentially small overlap (\ref{overlap}) of the replica and the vacuum states, such decays occur quite seldom. And indeed, according to Eq.~(\ref{w}), the probability for a macroscopically large domain to decay per unit time is exponentially suppressed.
One is therefore led to conclude that, if such encapsulated domains appeared in the Universe at its early
stages, they would have a chance to survive till the present time, remaining however dark to external observers.

\bigskip
The work of D.A. was supported by the Portuguese Foundation for Science and Technology
(FCT, program Ci\^encia-2008) and by 
the Center for Physics of Fundamental Interactions (CFIF) at Instituto Superior
T\'ecnico (IST), Lisbon. A.N. would like to thank the CFIF members for the warm hospitality extended to him during his
stay in Lisbon. The work of A.N. was supported by the
State Corporation of Russian Federation ``Rosatom'', by the FCT
(grant PTDC/FIS/70843/2006-Fi\-si\-ca), by the German Research Foundation (DFG, grant 436 RUS 113/991/0-1), by the
Russian Foundation for Basic Research (RFFI, grants
RFFI-09-02-91342-NNIOa and RFFI-09-02-00629a), as well as by the nonprofit Dynasty foundation and ICFPM.

\end{document}